\documentstyle[12pt]{article}
\newcommand\pubnumber{SLAC--PUB--95--6967}

\def\Title#1  {\begin{center}{\Large #1}\end{center}}
\def\Author#1 {\begin{center}{\sc #1}   \end{center}}
\def\Address#1{\begin{center}{\it #1}   \end{center}}
\def\doeack{\footnote{Work supported by the Department of Energy,
            contract DE--AC03--76SF00515.}}
\def\SLAC{Stanford Linear Accelerator Center\\
          Stanford University, Stanford, California 94309}
\newcommand\pubblock{\rightline{\begin{tabular}{l} \pubnumber\\
\end{tabular}}}
\newenvironment{Abstract} {\begin{quotation}
\begin{center}ABSTRACT\end{center}\bigskip}{\end{quotation}}
\def\Acknowledgements{\bigskip\bigskip {\begin{center}\begin{large}
\bf Acknowledgments \end{large}\end{center}}}
\def\etal{{\it et al.}}
\def\ie{{\it i.e.}}
\def\bar{\overline}
\def\longvec{\overrightarrow}
\def\rarrow{\rightarrow}
\def\cm{{\rm cm}}
\def\epem{{e^+e^-}}

\begin{document}
\begin{titlepage}
\pubblock
\vfill
\Title{Addendum to the Test of CP Violation in Tau Decay
\doeack}
\vfill
\Author{Yung Su Tsai}
\bigskip
\Address{\SLAC}
\vfill
\begin{Abstract}
We discuss the test of CP and CPT violation in $\tau$ decay without
using the polarized electron beam by comparing partial fractions
of $\tau^-$ and $\tau^+$ decay into channels with strong final state
interactions.  For example, $\Gamma(\tau^-\rarrow \pi^-+\pi^0+\nu)
\ne \Gamma(\tau^+\rarrow \pi^++\pi^0+\nu)$ signifies violation of
CP.  The optimum energy to investigatge CP violation in $\tau$ decay
is discussed.  We conclude that this energy is a few MeV below
$\psi(2s)$ in order to avoid the charm contribution and over
abundance of hadrons at the $\psi(2s)$ peak.
\end{Abstract}
\vfill

\begin{center}
Submitted to the Proceedings of the \\
Workshop on the Tau-Charm Factory \\
21--23 June 1995\\
Argonne National Laboratory, Argonne, Illinois
\end{center}
 \vfill

\end{titlepage}

\section{Introduction}

Understanding CP violation in the elementary particle system is a
fascinating subject in itself.  It is also a key to understanding
the preponderance of matter over antimatter in our universe.  Up to
now the only evidence of CP violation on the elementary particle
level is the decay of the $k_L$ system and this is too meager to
construct a credible standard theory for CP violation for all
particles.  In this paper we discuss measurement of CP violation in
$\tau$ decay.  This is an interesting subject because $\tau$ is the
heaviest lepton and thus if a charged Higgs boson is responsible for
CP violation we would most likely see the effect here among all the
leptons.  Also the Kobayashi-Maskawa theory \cite{refA} says that CP
violation should not occur in the leptonic sector because the gauge
eigenstate and mass eigenstate are identical in the lepton sector
due to zero neutrino masses in the Standard Model.  These basic
assumptions of Kobayashi-Maskawa must be tested.  CP violation in
$\tau$ has been investigated previously mainly in the production of
$\tau$ pair coming from the possible existence of the electric
dipole moment \cite{refB} of $\tau$.  However since the electric
dipole moment of $\tau$ is induced by weak or semiweak corrections
to the electromagnetic vertex of $\tau$ its effect is expected to
be less than $(m_\tau/m_w)^2\alpha = 3\times 10^{-6}$ and thus
impossible to detect even with $10^8$ $\tau$ pairs available in the
Tau-Charm Factory.  Similarly the interference between CP violating
neutral Higgs boson exchange and the one photon exchange diagrams is
also completely negligible \cite{refC,refD}.  Thus CP can be assumed
to be conserved in the production the $\tau$ pair; we need to
consider only CP violation in the decay of $\tau$.

Since the decay of $\tau$ is a weak interaction, if CP violation in
$\tau$ is weak, then its effect should be of order 1 whereas if it
is milliweak its effect should be of order $10^{-3}$ and detectable
with $10^8$ $\tau$ pairs available at the Tau-Charm Factory.

In my previous papers \cite{refC,refD} I have discussed how to use
the polarized electron beam to investigate CP violation in $\tau$
decay by constructing rotationally invariant quantities such as
$\longvec w_i\cdot \vec a, (\longvec w_i\times \vec a)\cdot \vec b,\
(\longvec w_i\times\vec\mu)\cdot\longvec w_\mu$ where $\vec a$ and
$\vec b$ are momenta of hadrons in the semileptonic decay of $\tau$;
$\vec \mu$ and $\longvec w_\mu$ are momentum and the polarizaiton of
muon in the decay $\tau^-\rarrow \mu^-+\nu_\tau+\bar\nu_\mu$ or its
charge conjugate; $\longvec w_i$ is the initial beam polarization
\begin{equation}
\longvec w_i = {w_1+w_2\over 1+w_1w_2}\, \hat e_z \ ,
\label{eq1}
\end{equation}
where $\hat e_z$ is the direction of the incident electron and $w_1$
and $w_2$ are polarization of the electron and positron in the $z$
direction.

In section 2 I point out that $\Gamma(\tau^-\rarrow \nu_\tau+a+b)
\ne \Gamma(\tau^+\rarrow \bar \nu_\tau+\bar a+\bar b)$ also
signifies CP violation.  We give the physical reason for it. We also
compare the merits of this kind of measurement with those using
polarized beams.  In section 3 we discuss the optimum energy to do
$\tau$ physics at the Tau-Charm Factory.

\section{CP Violation in $\tau$ Decay using Branching Fractions}

CPT conservation says that the total widths of $\tau^-$ and
$\tau^+$ must be equal.  Also partial widths into those channels
without final state interactions, such as $\tau^-\rarrow \mu^-+\bar
\nu_\mu+\nu_\tau$, $\tau^-\rarrow e^-+\bar\nu_e+\nu_\tau$,
$\tau^-\rarrow \pi^-+\nu_\tau$, and $\tau^-\rarrow k^-+\nu_\tau$,
must be the same as the corresponding channels for $\tau^+$ decay
\cite{refE}. However for decay channels that contain final state
interactions, such as $\tau^-\rarrow \pi^-+\pi^0+\nu_\tau$,
$\pi^-+\pi^-+\pi^++\nu_\tau$, $\pi^-+\pi^0+\pi^0+\nu_\tau$,
$\pi^-+k^0+\nu_\tau$, and $\pi^0+k^-+\nu_\tau$, the CP violation can
show up as the inequality in partial widths for charge conjugate
decay modes.  For example, $\Gamma(\tau^-\rarrow \pi^-+\pi^0+\nu)
\ne \Gamma(\tau^+\rarrow \pi^++\pi^0+\bar\nu)$ signifies violation
of CP, but $\Gamma(\tau^-\rarrow \mu^-+2\nu) \ne
\Gamma(\tau^+\rarrow \mu^++2\nu)$ or $\Gamma(\tau^-\rarrow
\mbox{all})\ne \Gamma(\tau^+\rarrow \mbox{all})$ will indicate that
CPT is violated. The polarization vector $\longvec w_i$ defined in
Eq. (\ref{eq1}) can be used to construct many rotationally invariant
products to investigate T, CP, CVC, and charged Higgs boson exchange
in leptonic \cite{refD} and semileptonic \cite{refC} decays of
$\tau$.  The polarization dependent quantities will yield
information on structure of CP violations whereas the polarization
independent quantities such as the difference in partial widths
between $\tau^-\rarrow \nu_\tau +\pi^-+\pi^0$ and $\tau^+\rarrow
\bar \nu_\tau+\pi^++\pi^0$ will merely indicate the existence and
magnitude of the CP violation. As pointed out in Ref. \cite{refC}
this difference in partial widths is due to the combined effects of
CP violation and the inelastic final state interaction such as
$2\pi$ going into $4\pi$ and vice versa. In the absence of CP
violation the probabilities of $2\pi$ going into $4\pi$ and vice
versa in the $\tau^-$ decay are equal to those in the $\tau^+$
decay.  However in the presence of CP violation the amplitudes for
the decay is proportional to $\exp(i\delta_w+i\delta_s)$ for
$\tau^-$ and $\exp(-i\delta_w+i\delta_s)$ for $\tau^+$ and thus they
become different.

\section{Optimum Energy to do $\tau$ Physics}

The energy of the machine should be set below charm threshold;
\ie\ $E_{\cm}=1869.3$ MeV for each beam.  Near the threshold of $\tau$
pair production, $\tau$ pair events can uniquely be identified by
$e$-hadron, $e$-$\mu$, $\mu$-hadron events.  Above the charm
threshold charm events produce the unwelcome leptonic background
\cite{refF}.  The best energy to run is either at $\psi(2s,\ 3,685$
MeV) or slightly below it.  The total cross section for $e^+e^-
\rarrow \psi(3.685) \rarrow \tau^+\tau^-$ can be written as
\cite{refG}
\begin{equation}
\sigma_r(w) = 12\pi\
       {\Gamma(\psi\rarrow 2e)\Gamma(\psi\rarrow2\tau)
        \over (w^2-M^2_R)^2+\Gamma^2_t M^2_R} \ ,
\label{eq2}
\end{equation}
where $w=2E$, $M_R=3.685$ GeV, $\Gamma_t = 243$ keV,
$\Gamma(\psi\rarrow 2e) = 21.4$ keV, and $\Gamma(\psi\rarrow 2\tau)
= \Gamma(\psi\rarrow 2e)(\beta(3-\beta^2)/2)$, with $\beta =
\sqrt{1-((2M_\tau)^2/w^2)} \approx 0.26426$, and $\beta(2-\beta^2)/2
\approx 0.38717$.  At the peak of the resonance we have
\begin{equation}
\sigma_r(M_R) = {12\pi\over M^2_R}\ B^2(\psi\rarrow 2e)\,
      0.38717 = 32.40 \times 10^{-33}\cm^2 \ .
\label{eq3}
\end{equation}
The peak cross section of $\epem \rarrow \tau^+\tau^-$ at continuum
which occurs at $2E= 4.174$ GeV is \cite{refC,refD}
\begin{equation}
\sigma_c(4.174) = {\pi\alpha^2\over 6}\times 1.036\ {1\over
      M^2_\tau} = 3.562\times 10^{-33}\cm^2 \ .
\label{eq4}
\end{equation}
Thus
\begin{equation}
{\sigma_r(3.685)\over \sigma_c(4.174)} = 9.096 \ .
\label{eq5}
\end{equation}
This number must be reduced because the machine width is much wider
than the resonance width and the radiative corrections further
broaden the effective machine width.  This problem was first solved
\cite{refG} by the author in 1974 immediately after the discovery of
$J/\psi$.  The most comprehensive account was given in Ref.
\cite{refH} which we follow here.  Qualitatively if the machine width
is $\Delta$ and the resonance width is $\Gamma_t$, then only the
fraction $\Gamma_t/\Delta$ of the beam is effective in producing the
resonance peak if $\Delta \gg \Gamma_t$.  The effect of radiative
corrections can be estimated by the change in the height of the
Gaussian peak of the machine energy by the radiative corrections
because only the peak height matters when the resonance is narrower
than the beam width.  The result is \cite{refH}
\begin{equation}
\sigma_{\exp}(3.685) = \sigma_r(3.685)\
\left[\sqrt{{\pi\over8}}\ {\Gamma_t\over\Delta}\right]
\left[\left({\sqrt8\Delta\over
3.685}\right)^T\Gamma\left({T\over2}+1\right)\right] +
\sigma_c(3.685)
\label{eq6}
\end{equation}
where $\sigma_c(3.685) = 2.476\times 10^{-33}\cm$ is the continuum
cross section, $\Delta$ is the Gaussian beam width defined by
\begin{equation}
G(w,w^\prime) = {1\over\sqrt{2\pi}\Delta}
\exp\left[-{(w-w^\prime)^2\over 2\Delta^2}\right]\ ,
\label{eq7}
\end{equation}
and is related to the full width at a half maximum (FWHM) by
\begin{equation}
\Delta = {(FWHM) \over 2.3848} \ .
\label{eq8}
\end{equation}
$T$ is called the equivalent radiator thickness defined by
\begin{equation}
T = {2\alpha\over\pi}\
    \left[\ell n\ {M^2_R\over m^2_e}-1\right] = 0.14229 \ .
\label{eq9}
\end{equation}
$\Gamma$ is the Gamma function and its value is
\begin{equation}
\Gamma \left(1+{T\over 2}\right) = 0.96365 \ .
\label{eq10}
\end{equation}
The first square bracket shows that only a fraction of the incoming
beam, $\Gamma_t/\Delta$, is effective in producing the resonance.
The factor $\sqrt{\pi/8}$ comes form the fact that $\Gamma_t$ is the
FWHM of the Breit-Wigner formula whereas FWHM of the Gaussian beam
profile is given by Eq. (\ref{eq8}).  The second square bracket
represents the reduction of the Gaussian peak height due to the
photon emission whose effective cutoff is $\Delta E = \sqrt8\,
\Delta$.  The Gamma function, Eq. (\ref{eq10}), comes from the
folding of the Gaussian function with the photon straggling function
\cite{refH}.  At the Beijing Electron-Positron Collider $\Delta =
1.4$ MeV and thus from Eq. (\ref{eq6}) we have
\begin{equation}
\sigma_{\exp}(2.685,\Delta=1.4\, \mbox{MeV}) = 0.0411\,
\sigma_r(3.685)+\sigma_c(3.685) \ .
\label{eq11}
\end{equation}
there is a scheme \cite{refI} to make $\Delta$ as small as 0.14 MeV
using a monochromatizer; we have then
\begin{equation}
\sigma_{\exp}(3.685,\Delta=0.14\, \mbox{MeV}) = 0.286\,
\sigma_r(3.685)+\sigma_c(3.685) \ .
\label{eq12}
\end{equation}
Since the branching fraction to $\tau$ pair is 0.34\%\ in
$\sigma_r(3.685)$ there are several hundred $\pi$'s for each
$\tau$ pair produced by $\sigma_r(3.685)$.

The BES Collaboration \cite{refF} has successfully carried out
$\tau$ experiments using $\psi^\prime$ under the conditions shown in
Eq. (\ref{eq11}), where the first term is about 0.48 of the last
term. For their experiment the hadron background did not cause any
problem for four reasons: (1) most of the hadron backgrounds are
multiprong events whereas $\tau$ events are mostly two-prong events.
This fact can be used to eliminate the background. (2) They did not
use the monochromatizer. (3) Particle ID has about $10^{-3}$
efficiency. (4) Accuracy of $10^{-2}$ is good enough for them,
whereas CP experiment needs $10^{-3}$ accuracy.

An alternative to use Eq. (\ref{eq11}) or (\ref{eq12}) is to avoid
$\psi^\prime$ all together and run the machine at a slightly lower
energy, say at 3.680 GeV.  From the consideration of background this
is probably the ideal energy to run the Tau-Charm Factory. At
$W=3.680$ GeV  the component of polarization of $\tau^\pm$ in the
beam direction averaged over the production angle is slightly
improved:
\[
\bar w_z = \int^1_{-1}w_z\, {d\sigma\over d\cos\theta}\,
     d\cos\theta\bigg/\sigma = {w_1+w_2\over 1+w_1w_2}\
     {1+2a\over 2+a^2}\equiv {w_1+w_2\over 1+w_1w_2}\ F(a) \ ,
\]
where $a=2M_\tau/W$.  At $w=4.174$ GeV we have $F= 0.992$, but at
$w=3.680$ GeV we have $F=0.9996$. The cross section is reduced from
$\sigma_c(4.174) = 3.562\times 10^{-33}\cm^2$ to $\sigma_c(3.680) =
2.44\times 10^{-33}\cm^2$.  This energy is preferred in order to
avoid both the charm background and overabundance of hadrons in the
$\psi(2s)$ peak.

\Acknowledgements
The author wishes to thank Bill Dunwoodie, W.K.H. Panofsky, and
Martin Perl  for discussions on the optimum energy for doing $\tau$
physics at the Tau-Charm Factory. I would also like to thank Karl
Brown for explaining to me how the beam monochromatizer works.

\end{document}